\documentclass[twocolumn]{aa}
\usepackage{graphicx}
\usepackage{txfonts}
\usepackage{lscape}
\begin{document}

\title{Photometric properties and origin of bulges in SB0 galaxies}

\author{J. A. L. Aguerri\inst{1}, N. Elias-Rosa\inst{2}, E. M. Corsini\inst{3}, and 
C. Mu\~noz-Tu\~n\'on\inst{1}}

\offprints{jalfonso@ll.iac.es}

\institute{Instituto de Astrof\'\i sica de Canarias, Calle Via Lactea s/n, 
  E-38200 La Laguna, Spain
\and INAF--Osservatorio Astronomico di Padova, vicolo dell'Osservatorio 5, 
  I-35122 Padova, Italy
\and Dipartimento di Astronomia, Universit\`a di Padova, 
vicolo dell'Osservatorio 2, I-35122 Padova, Italy}

\date{\today}

\authorrunning{Aguerri et al.}
\titlerunning{Bulges of SB0 galaxies}

\abstract{
We have derived the photometric parameters for the structural
components of a sample of fourteen SB0 galaxies by applying a parametric
photometric decomposition to their observed $I$-band surface
brightness distribution.
We find that SB0 bulges are similar to bulges of the early-type
unbarred spirals, i.e. they have nearly exponential surface
brightness profiles ($\left<n\right>=1.48\pm0.16$) and their effective
radii are strongly coupled to the scale lengths of their surrounding
discs ($\left<r_{\rm e}/h\right>=0.20\pm0.01$). The photometric
analysis alone does not allow us to differentiate SB0 bulges from
unbarred S0 ones. 
However, three sample bulges have  disc properties typical of
pseudobulges. The bulges of NGC 1308 and NGC 4340 rotate faster than
bulges of unbarred galaxies and models of isotropic oblate spheroids
with equal ellipticity. The bulge of IC 874 has a velocity dispersion
lower than  expected from the Faber--Jackson correlation and the
fundamental plane of the elliptical galaxies and S0 bulges. The
remaining sample bulges are classical bulges, and are kinematically
similar to lower-luminosity ellipticals. In particular, they follow
the Faber--Jackson correlation, lie on the fundamental plane and those
for which stellar kinematics are available rotate as fast as the bulges
of unbarred galaxies.
\keywords{galaxies: bulge -- 
galaxies: elliptical and lenticular, cD --
galaxies: fundamental parameters --
galaxies: photometry --
galaxies: structure}}

\maketitle


\section{Introduction}

Stellar bars are fairly ubiquitous in galactic discs. They have been
found in a large fraction of spiral galaxies (Eskridge et al.\ 2000)
and in almost all Magellanic-type galaxies (Odewahn 1996). The
presence of a bar affects the morphological, dynamical and
chemical evolution of the host galaxy.

In particular, gas flows and the secular evolution of barred galaxies
have been claimed to be viable mechanisms leading to the formation
and/or growth of the bulge and have been studied by means of numerical
simulation for a long time (see Friedli 1999 for a review).

Numerical simulations have shown that gravitational torques from bars
are very efficient at driving interstellar gas toward galactic
centres.  The main consequences of this gas accumulation is the
triggering of bursts of circumnuclear star formation and the formation
of a central mass concentration, such as compact bulges, which 
may weaken or dissolve the bar (Pfenniger \& Norman 1990; Hasan \&
Norman 1990). The efficiency of the bar destruction depends on both
the concentration of the central component and its growth rate (e.g.,
Bournaud \& Combes 2002; Shen \& Sellwood 2004). Besides bar-driven
accretion, external processes, including satellite accretion
(Pfenniger 1991; Aguerri et al.\ 2001) and galaxy merging (Barnes \&
Hernquist 1991), are also able to build a central  mass reservoir
which makes the bulge grow and the bar dissolve.

Moreover, according to the results of $N$-body simulations, the inner
parts of a bar inflate after a few bar rotations because of
large-scale violent bending instabilities and settle with an increased
thickness and vertical velocity dispersion (e.g., Combes \& Sanders
1981; Combes et al.\ 1990; Raha et al.\ 1991). This leads to the
establishment of the connection between the bar-buckling mechanism and
the formation of boxy/peanut bulges (Bureau \& Freeman 1999; Bureau \&
Athanassoula 1999; Chung \& Bureau 2004). The buckling
instability does not destroy the bar and forms a central stellar
condensation reminiscent of the bulges of late-type spirals
(Debattista et al.\ 2004). This result is in agreement with the
early findings of Hohl (1971).

It has been inferred that all the above processes increase the
bulge-to-disc ratio, driving the evolution of spiral galaxies from
late to early Hubble types (Friedli \& Martinet 1993; Hasan et al.\
1993). The aim of this paper is to perform a systematic study of the
photometric properties of bulges in early-type barred galaxies. To
date, this crucial piece of information is still missing. Few barred
galaxies have been studied in such detail (e.g., Prieto et al.\ 1997,
2001) because of the complication introduced into the photometric
analysis by the presence of the bar component.
For example, there are claims that bars in early- and late-type disc
galaxies have flat and exponential surface brightness profiles,
respectively (Elmegreen et al.\ 1996), although the S\'ersic profile
(S\'ersic 1968) has been adopted too (Peng et al.\ 2002).

By properly taking into account the bar contribution to the galaxy
surface brightness we derive the structural parameters of SB0 bulges.
To this purpose, we give a purely photometric and parametric
definition of bulges following Fathi \& Peletier (2003). In
particular, we assume that they are the inner photometric component of
our sample galaxies, and that their surface brightness profile is
described by the S\'ersic law.
The comparison with bulges in unbarred S0 and spiral galaxies will
provide an insight as to whether SB0 bulges are bar-derived,
disc-like, or otherwise different from classical early-type bulges
(Kormendy 1982, 1993; Kormendy \& Illingworth 1983; M\"ollenhoff et
al. 1995; Seifert \& Scorza 1996; Erwin et al. 2003; Balcells et
al. 2003).

This paper is organized as follows. In Sect. \ref{sec:sample} we
present an overview of the properties of the sample galaxies as well
as their photometric observations and data analysis.
In Sect. \ref{sec:decomposition} we derive the photometric parameters
of the structural components (bulge, disc, bar and lens) of the sample
galaxies by applying a parametric photometric decomposition to their
observed $I$-band surface brightness distribution.
In Sect. \ref{sec:discussion} we compare the structural parameters of
SB0 bars and bulges to those obtained in numerical simulations of
early-type barred galaxies and to those derived for bulges of unbarred
galaxies, respectively. Moreover, we derive the location of SB0 bulges
on the fundamental plane and Faber--Jackson relation of elliptical
galaxies and S0 bulges by means of the central stellar velocity
dispersion available for thirteen sample galaxies.
In Sect. \ref{sec:conclusions} we discuss a possible scenario for the
formation of SB0 bulges.

\section{Sample selection, observations and basic data reduction}
\label{sec:sample}

We started a programme some years ago to enlarge the sample of
early-type barred galaxies with pattern speeds measured using the
Tremaine--Weinberg method (Tremaine \& Weinberg 1984), including
galaxies of various bar strengths, luminosities, inclinations, etc. To
this end, we selected a sample of fourteen SB0 objects by a visual
inspection of their images in the Digitized Sky Survey. The galaxies
were chosen for their brightness and undisturbed morphology, with no
evidence of dust or spiral arms to complicate the Tremaine-Weinberg
analysis, and a bar at an angle intermediate between the orientation
of the major and minor axes. To date, we have derived the bar pattern
speed of seven objects, namely NGC 1023 (Debattista et al.\ 2002), NGC
2950 (Corsini et al.\ 2003), ESO 139$-$G009, IC 874, NGC 1308, NGC
1440 and NGC 3412 (Aguerri et al.\ 2003). In this paper, we focus on
the $I$-band photometric properties of the structural components of
all the galaxies in the sample except for NGC 1023. Its high
inclination (Debattista et al.\ 2002) prevents a reliable photometric
decomposition from being obtained. In addition, we considered the SB0
galaxy NGC 7079, for which the bar pattern speed has been recently
obtained by Debattista \& Williams (2004), who made their $I$-band
images available to us. A compilation of the main properties of the
final collection of fourteen early-type barred galaxies is given in
Table \ref{tab:sample}.

The photometric observations of sample galaxies were carried out
during several observing runs between 1997 and 2003 at the Cerro
Tololo Inter-American Observatory (CTIO, Chile), at the European
Southern Observatory (ESO) in La Silla (Chile), and at the Roque de
los Muchachos Observatory (ORM) on La Palma (Spain). The log of
observations and details of the different instrumental setups are
given in Table \ref{tab:log}.

All images were reduced using standard IRAF\footnote{IRAF is
distributed by NOAO, which is operated by AURA Inc., under contract
with the National Science Foundation.} tasks. We first subtracted a
bias frame consisting of ten exposures for each night. The images were
flat-fielded using sky flats taken at the beginning and/or end of each
observing night. The sky background level was removed by fitting a
second-order polynomial to the regions free of sources in the
images. Special care was taken during sky subtraction to reach the
outermost parts of the objects. Cosmic rays were removed by combining
the different exposures using field stars as a reference and adopting
a sigma clipping rejection algorithm. For the photometric calibration
of the galaxies, standard stars of known magnitudes were observed. The
calibration constant includes corrections for atmospheric and Galactic
extinction, and a colour term. No attempt was made to correct for
internal extinction. The atmospheric extinction was taken from the
differential aerosol extinction for CTIO (Baldwin
\& Stone 1984), ESO (Burki et al.\ 1995) and the ORM (King 1985). The
Galactic extinction in the $B$ band was taken from Schlegel et al.\
(1998). We used the Galactic extinction law from Cardelli et al.\
(1989) in order to get the extinction in the $I$ band. Figure
\ref{fig:contoura} shows the calibrated $I$-band images of the sample
galaxies.

\begin{table*}
\begin{flushleft} 
\caption[]{Parameters of the sample galaxies}
\label{tab:sample} 
\begin{tabular}{llcrcrrccrc}
\hline
\noalign{\smallskip}
\multicolumn{1}{c}{Galaxy} &
\multicolumn{1}{c}{Type} &
\multicolumn{1}{c}{$i$} &
\multicolumn{1}{c}{PA} &
\multicolumn{1}{c}{$B_T$} &
\multicolumn{1}{c}{$D_{25}\times d_{25}$} &
\multicolumn{1}{c}{$V_{\rm CMB}$} &
\multicolumn{1}{c}{$D$} &
\multicolumn{1}{c}{$M_{B_T}^0$} &
\multicolumn{1}{c}{$\sigma_0$} &
\multicolumn{1}{c}{Ref.} \\
\multicolumn{1}{c}{} &
\multicolumn{1}{c}{(RC3)} &
\multicolumn{1}{c}{($^\circ$)} &
\multicolumn{1}{c}{($^\circ$)} &
\multicolumn{1}{c}{(mag)} &
\multicolumn{1}{c}{(arcsec)} &
\multicolumn{1}{c}{(km s$^{-1}$)} &
\multicolumn{1}{c}{(Mpc)} &
\multicolumn{1}{c}{(mag)} &
\multicolumn{1}{c}{(km s$^{-1}$)} &
\multicolumn{1}{c}{} \\
\multicolumn{1}{c}{(1)} &
\multicolumn{1}{c}{(2)} &
\multicolumn{1}{c}{(3)} &
\multicolumn{1}{c}{(4)} &
\multicolumn{1}{c}{(5)} &
\multicolumn{1}{c}{(6)} &
\multicolumn{1}{c}{(7)} &
\multicolumn{1}{c}{(8)} &
\multicolumn{1}{c}{(9)} &
\multicolumn{1}{c}{(10)} &
\multicolumn{1}{c}{(11)} \\
\noalign{\smallskip}
\hline
\noalign{\smallskip}
ESO 139$-$G009  & (R)SAB0(rs)  &  46 &  94 & 14.35 &   $74\times60$& 5389 & 71.9 & $-20.28$ & 196 & 1\\
IC 874        & SB0(rs)      &  39 &  27 & 13.60 &   $60\times43$  & 2602 & 34.7 & $-19.41$ &  99 & 1\\
IC 4796       & SB0(s):      &  55 & 126 & 13.25 &   $95\times56$ & 2983 & 39.8 & $-20.13$ & 143 & 2\\
NGC 357       & SB0(r):      &  36 &  20 & 13.12 & $144\times104$ & 2136 & 28.5 & $-19.78$ & 178 & 2\\
NGC 364       & (R)SB0(s):   &  46 &  27 & 15.61 &   $85\times76$ & 4883 & 65.1 & $-18.68$ & ... & ...\\
NGC 936       & SB0$^+$(rs)  &  49 & 136 & 11.16 & $281\times244$ & 1100 & 14.7 & $-19.85$ & 204 & 2\\
NGC 1308      & SB0(r)       &  36 &  59 & 14.70 &   $70\times51$ & 6180 & 82.4 & $-19.88$ & 219 & 1\\
NGC 1440      &(R$'$)SB0(rs):&  38 &  26 & 12.90 &  $128\times97$ & 1382 & 18.4 & $-18.90$ & 195 & 1\\
NGC 2950      & (R)SB0(r)    &  46 & 116 & 11.82 & $161\times107$ & 1510 & 20.1 & $-19.79$ & 164 & 3\\
NGC 3412      & SB0(s)       &  55 & 151 & 11.43 & $218\times123$ & 1202 & 16.0 & $-19.73$ & 112 & 1\\
NGC 3941      & SB0(s)       &  49 &   9 & 11.23 & $208\times137$ & 1183 & 15.8 & $-19.87$ & 137 & 4\\
NGC 4340      & SB0$^+$(r)   &  49 &  95 & 12.11 & $213\times169$ & 1238 & 16.5 & $-19.11$ & 118 & 5\\  
NGC 6684      & (R$'$)SB0(s) &  49 &  36 & 11.29 & $239\times158$ &  837 & 11.2 & $-19.26$ & 112 & 2\\
NGC 7079      & SB0(s)       &  52 &  82 & 12.49 &  $128\times79$ & 2541 & 33.9 & $-20.34$ & 165 & 2\\
\noalign{\smallskip}
\hline
\noalign{\bigskip}
\end{tabular}
\begin{minipage}{17cm}
  NOTE. Col.(2): morphological classification from de Vaucouleurs et
  al.\ (1991, hereafter RC3); Col.(3): inclination of the galaxy disc
  from this paper; Col.(4): major-axis position angle of the galaxy
  disc from this paper; Col.(5): total observed blue magnitude from
  from Lyon Extragalactic Database (hereafter LEDA); Col.(6): apparent
  isophotal diameters measured at a surface-brightness level of $\mu_B
  = 25$ mag arcsec$^{-2}$ from RC3; Col.(7): radial velocity with
  respect to the CMB radiation from LEDA; Col.(8): distance obtained
  as $V_{\rm CMB}/H_0$ with $H_0 = 75$ km s$^{-1}$ Mpc$^{-1}$;
  Col.(9): absolute total blue magnitude from $B_T$ corrected for
  inclination and extinction as in LEDA and adopting $D$; Col.(10):
  central velocity dispersion of the stellar component after
  correcting to the equivalent of an aperture of radius $r_{\rm e}/8$
  following the prescription by J\/orgensen et al.\ (1995), with
  $r_{\rm e}$ the effective radius of the bulge from Table
  \ref{tab:parameters}.   Typical error is 10 km s$^{-1}$.
  Col.(11): list of references for the central velocity dispersion:
  1=Aguerri et al. (2003), 2=Wegner et al.\ (2003), 3=Corsini et al.\
  (2003), 4=Fisher (1997), 5=Prugniel \& Simien (1997).
\end{minipage}
\end{flushleft} 
\end{table*}

\begin{table*}
\begin{flushleft} 
\caption[]{Log of the observations}
\label{tab:log} 
\begin{tabular}{lllcclrcc}
\hline
\noalign{\smallskip}
\multicolumn{1}{c}{Galaxy} &
\multicolumn{1}{c}{Instrument} &
\multicolumn{1}{c}{CCD} &
\multicolumn{1}{c}{Pixel \#} &
\multicolumn{1}{c}{Scale} &
\multicolumn{1}{c}{Date} &
\multicolumn{1}{c}{Exp.\ time} &
\multicolumn{1}{c}{Seeing} &
\multicolumn{1}{c}{$\mu_I$} \\
\multicolumn{1}{c}{} &
\multicolumn{1}{c}{} &
\multicolumn{1}{c}{} &
\multicolumn{1}{c}{} &
\multicolumn{1}{c}{($''$/pixel)} &
\multicolumn{1}{c}{} &
\multicolumn{1}{c}{(s)} &
\multicolumn{1}{c}{($''$)} &
\multicolumn{1}{c}{(mag/arcsec$^{2}$)} \\
\multicolumn{1}{c}{(1)} &
\multicolumn{1}{c}{(2)} &
\multicolumn{1}{c}{(3)} &
\multicolumn{1}{c}{(4)} &
\multicolumn{1}{c}{(5)} &
\multicolumn{1}{c}{(6)} &
\multicolumn{1}{c}{(7)} &
\multicolumn{1}{c}{(8)} &
\multicolumn{1}{c}{(9)} \\
\noalign{\smallskip}
\hline
\noalign{\smallskip}	
ESO 139$-$G009 & NTT$+$EMMI& TK2048 EB & $2048\times4096$        & 0.270 & 2001 May 23 &  300$^g$ & 1.2 &22.05\\
IC 874   & NTT$+$EMMI    & TK2048 EB & $2048\times4096$        & 0.270 & 2001 May 23 &  300$^g$ & 0.9 &23.00\\
IC 4796  & Danish$+$DFOSC& MAT/EEV   & $2048\times4096$        & 0.390 & 2002 May 08 & 1260$^g$ & 1.1 &23.33\\
NGC 357  & NTT$+$EMMI    & MIT/LL    & $2\times2048\times4096$ & 0.332 & 2002 Oct 05 &  300$^g$ & 1.0 &24.06\\
NGC 364  & Danish$+$DFOSC& MAT/EEV   & $2048\times4096$        & 0.390 & 2001 Nov 11  &  600$^g$ & 1.2 &23.25\\
NGC 936  & JKT           & SITe2     & $2048\times2048$        & 0.330 & 2002 Sep 09 & 1800$^h$ & 1.5 &22.80\\
NGC 1308 & JKT           & SITe2     & $2048\times2048$        & 0.330 & 2001 Oct 11 & 7200$^h$ & 0.9 &22.20\\
NGC 1440 & Danish$+$DFOSC& MAT/EEV   & $2048\times4096$        & 0.390 & 2001 Nov 11 & 3600$^g$ & 1.1 &22.53\\
NGC 2950 & JKT           & SITe2     & $2048\times2048$        & 0.330 & 2000 Dec 28 & 2700$^h$ & 1.0 &23.15\\
NGC 3412 & JKT           & SITe2     & $2048\times2048$        & 0.330 & 2000 May 28 & 3600$^h$ & 1.0 &23.22\\
NGC 3941 & TNG$+$DOLORES & Loral     & $2048\times2048$        & 0.275 & 2003 Mar 11 &  120$^J$ & 1.1 &23.00\\
NGC 4340 & TNG$+$DOLORES & Loral     & $2048\times2048$        & 0.275 & 2003 Mar 11 &  180$^j$ & 1.0 &23.60\\
NGC 6684 & NTT$+$EMMI    & MIT/LL    & $2\times2048\times4096$ & 0.332 & 2002 Oct 05 &   75$^g$ & 1.2 &22.28\\
NGC 7079 & CTIO 0.9-m    & Tek 2K\#3 & $2048\times2046$        & 0.396 & 1997 Aug 02 &  900$^g$ & 1.5 &23.80\\
\noalign{\smallskip}
\hline
\noalign{\bigskip}
\end{tabular}
\begin{minipage}{17cm}
  NOTE. Col.(2): telescope and camera, NTT$+$EMMI = New Technology
  Telescope mounting the ESO Multi-Mode Instrument in Red Imaging and
  Low-Dispersion Spectroscopic Mode, Danish$+$DFOSC = 1.54 m Danish
  telescope mounting the Danish Faint Object Spectrograph and Camera,
  JKT = Jacobus Kapteyn Telescope, TNG$+$DOLORES = Telescopio
  Nazionale Galileo mounting the Device Optimized for the Low
  Resolution, CTIO 0.9-m = 0.9 m telescope at CTIO; Col.(7): total
  exposure time obtained with with Gunn ($^g$), Harris ($^h$) or
  Johnson ($^j$) $I-$band filter; Col.(8): seeing FWHM measured by
  fitting a circular two-dimensional Gaussian to several field stars
  in the final combined image of the galaxy; Col.(9): surface
  brightness level at the outermost radius along the bar major axis.
\end{minipage} 
\end{flushleft} 
\end{table*}

\begin{figure*}
\centering
\includegraphics[angle=0,width=18cm]{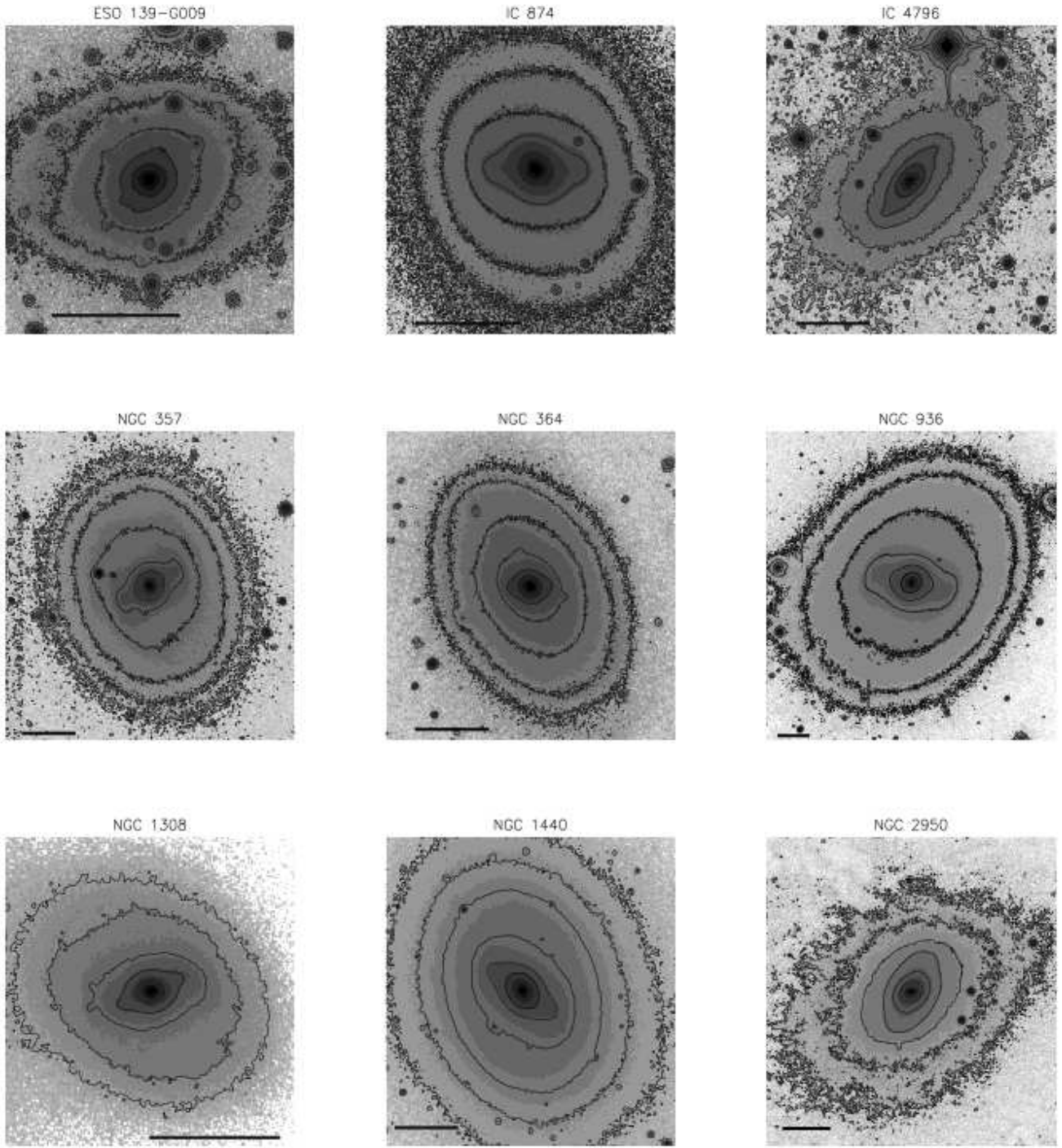}
\caption{$I$-band images of the galaxies: IC 4796, NGC 357, NGC 364,
   NGC 936, NGC 2950, NGC 3941, NGC 4340, NGC 6684 and NGC 7079.  In each panel north
   is at the top and east is to the left, and the bottom horizontal line is 30
   arcsec in length.  The isocontours are spaced at 1.0 mag arcsec$^{-2}$, with
   the outermost contour corresponding to the surface brightness at
   column 9 of Table 2.}   
\label{fig:contoura} 
\end{figure*}

\addtocounter{figure}{-1}

\begin{figure*}
\centering
\includegraphics[angle=0,width=18cm]{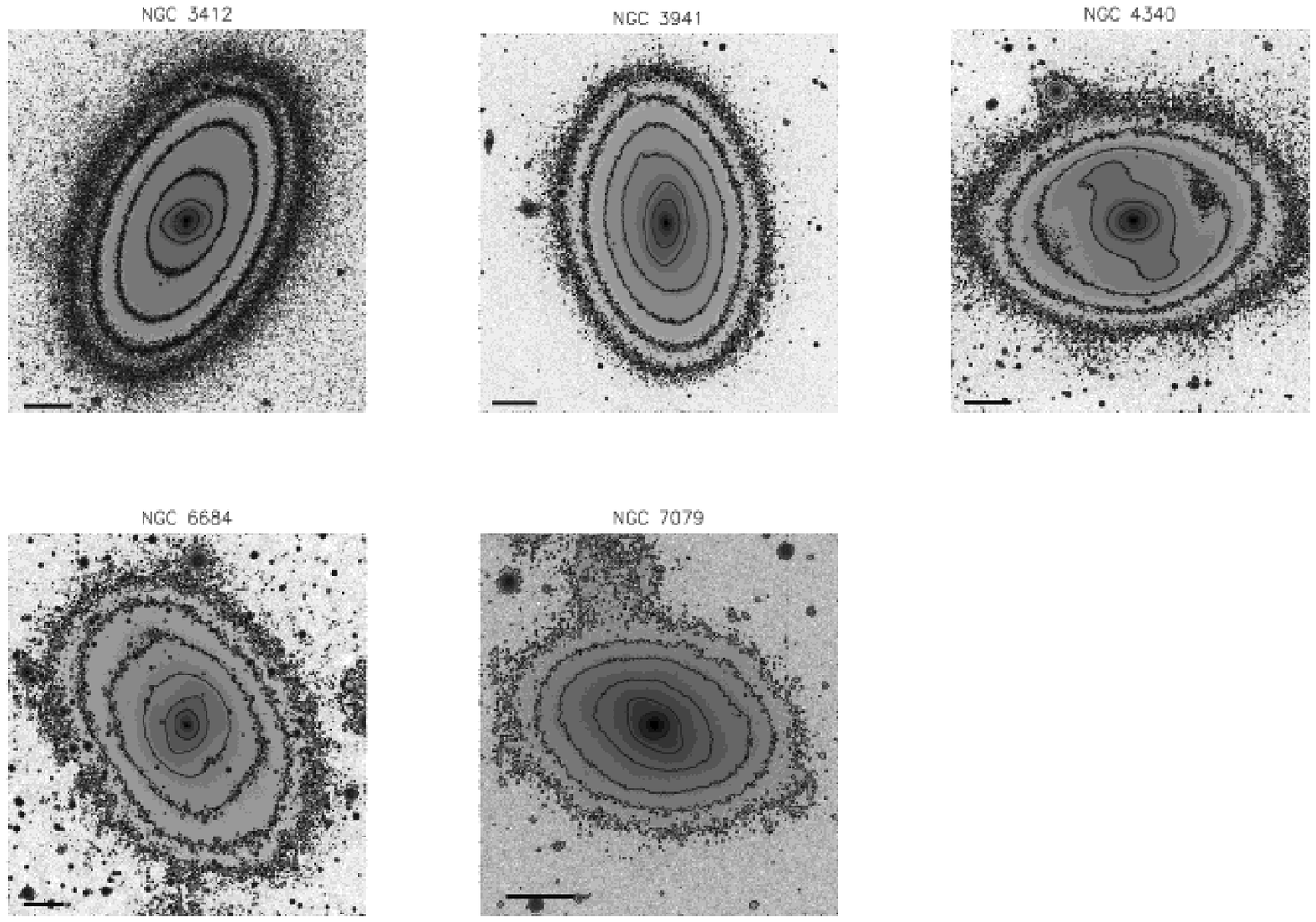}
\caption{(continued)}  
\label{fig:contourb} 
\end{figure*}

\section{Photometric parameters of the galaxy structural components}
\label{sec:decomposition}

\subsection{Decomposition of the radial surface brightness  profiles}
\label{sec:decomposition_lens}

We modeled the $I$-band surface brightness distribution of the sample
galaxies as the sum of the contribution of their structural components
(i.e., bulge, disc, bar and possibly a lens) by adopting the
photometric decomposition technique developed by Prieto et al.\
(2001).
The surface brightness distribution of each individual component has
been assumed to follow a parametric law, which has to be strictly
considered  as an empirical fitting function.

For the radial surface brightness profile of the bulge we assumed the
S\'ersic law,
\begin{eqnarray}
I_{\rm bulge}(r)=I_{\rm e}10^{-b_{n}\left[\left(r/r_{\rm
e}\right)^{1/n}-1\right]},
\end{eqnarray}
\noindent
where $r_{\rm e}$, $I_{\rm e}$ and $n$ are the effective (or
half-light) radius, the surface brightness at $r_{\rm e}$, and a shape
parameter describing the curvature of the profile, respectively. The
value of $b_n$ is coupled to $n$ so that half of the total flux is
always within $r_{\rm e}$ and can be approximated as
$b_{n}=0.868n-0.142$ (Caon et al.\ 1993).

For the radial surface brightness profile of the disc we assumed the
exponential law (Freeman 1970)
\begin{equation}
I_{\rm disc}(r)=I_0 e^{-r/h},
\end{equation}
\noindent
where $h$ and $I_{0}$ are the scale length and central surface
brightness of the disc, respectively.

For the radial surface brightness profile of the bar we assumed either
a flat profile (Prieto et al.\ 1997),
\begin{equation}
I_{\rm bar}(r)=I_b/\left(1+e^{\frac{r-r_{\rm b}}{r_{\rm s}}}\right),
\end{equation}
or a  Freeman bar profile (Freeman 1966),
\begin{equation}
I_{\rm bar}(r)=I_{\rm b}\sqrt{1-\left(\frac{r}{r_{\rm b}}\right)^{2}},
\end{equation}
\noindent
where $r_{\rm b}$, $r_{\rm s}$ and $I_{\rm b}$ are the bar length, a
scale length, and the bar central surface brightness, respectively.

For the radial surface brightness profile of the lens we assumed the
law by Duval \& Athanassoula (1983),
\begin{equation}
I_{\rm lens}(r)=I_{\rm l}\left(1-\left(\frac{r}{r_{\rm
l}}\right)^{2}\right),
\end{equation}
\noindent
where $r_{\rm l}$ and $I_{\rm l}$ are the length and central surface
brightness of the  lens.

Following Prieto et al.\ (2001), we extracted the radial surface
brightness profiles for each sample object along the major axis of the bar from the deprojected image of each galaxy. Image
deprojection was performed by means of a flux-conserving stretch
of the observed image along the disc minor axis, and adopting the disc
inclination and position angle given in Table \ref{tab:sample}.  They
were measured by averaging the values in the outermost region of the
ellipticity and position angle radial profiles, which were obtained by
fitting elliptical isophotes with the ELLIPSE task in IRAF. In all
cases, we first masked foreground stars and bad pixels and then we
fitted ellipses allowing their centres to vary. Within the errors, no
variation in the ellipse centres was found for the galaxies studied in
this paper. The final ellipse fits were done at fixed ellipse
centres. The inclination and position angle of the galaxies were
determined by averaging the outer isophotes, and their typical errors
are $\la1^{\circ}$.

Iterative fitting of the model surface brightness $I_{\rm mod}= I_{\rm
bulge} + I_{\rm disc} + I_{\rm bar} + I_{\rm lens}$ to the extracted
radial surface brightness profiles resulted in a non-linear system of
equations for the free parameters. These ranged from 7 (when bulge,
disc and Freeman bar were fitted) to 12 (when bulge, disc, flat bar
and lens were fitted). A Levenberg--Marquardt algorithm (Press et al.\
1992) was used for the solution of this system.
We first fitted the radial surface brightness profiles by assuming the
bulge shape parameter to be $n=0.5,1,1.5,$\ldots$,5$.  The result
corresponding to the lowest value of $\chi^2$ yielded a rough guess
for the initial values of the parameters in the Levenberg--Marquardt
iteration, where all the parameters, including $n$, were allowed to
vary. This approach ensures a robust estimate of $n$ and is similar to
the one extensively tested by MacArthur et al.\ (2003).

 The fit of the observed surface brightness profiles was done to
intensities. Errors in intensities were used in the fit. For all the
sample galaxies the best fit of the observed surface brightness was
obtained with a flat bar, except for NGC 3412 and NGC 6684, for which
a  Freeman bar was adopted.
A lens component was included to account for the light bump observed
at large radii in the radial surface brightness profiles of ESO
139$-$G009, IC 874, NGC 364, NGC 936, NGC 1440, NGC 3412 and NGC 7079.
Although it decreases the fit residuals, the choice of such a
lens component is somewhat arbitrary and alternative solutions like a
ring (e.g., Duval \& Athanassoula 1983) or a truncated disk (e.g.,
Pohlen et al. 2002) have not been considered.  For this reason, we
will discuss how the photometric parameters of the remaining
structural components and their relations are affected by the presence
of the lens component.
On the other hand, we did not consider as further components
the secondary stellar bars hosted by NGC 357, NGC 2950, NGC 3941, NGC
4340, and NGC 6684.  These secondary bars have a small size
($\simeq6''$) and are not aligned with the major axis of the main bar
(Erwin 2004 and references therein). This means that their
contribution to the surface brightness profile observed along the main
bar major axis is negligible and does not affect the bulge parameters
we obtained from such profiles.

The parameters derived for the structural components of the sample
galaxies are collected in Table \ref{tab:parameters}.  The formal
errors of the fit are given in Table \ref{tab:parameters} too. They
are most probably underestimated since they do not include the other
uncertainties, like those in the sky background determination. The
result of the photometric decomposition of the radial surface
brightness profiles extracted along the bar major axis of the sample
galaxies is shown in Figure
\ref{fig:decomposition}.

\begin{landscape}
\begin{table}
\caption[]{ Photometric parameters of the sample galaxies}
 \label{tab:parameters} 
\begin{tabular}{lcccccccccc}
\hline
\noalign{\smallskip}
\multicolumn{1}{c}{Galaxy} &
\multicolumn{1}{c}{$\mu_{\rm e}$} &
\multicolumn{1}{c}{$r_{\rm e}$} &
\multicolumn{1}{c}{$n$} &
\multicolumn{1}{c}{$\mu_0$} &
\multicolumn{1}{c}{$h$} &
\multicolumn{1}{c}{$\mu_{\rm b}$} &
\multicolumn{1}{c}{$r_{\rm b}$} &
\multicolumn{1}{c}{$r_{\rm s}$} &
\multicolumn{1}{c}{$\mu_{\rm l}$} &
\multicolumn{1}{c}{$r_{\rm l}$} \\
\multicolumn{1}{c}{ } &
\multicolumn{1}{c}{(mag/arcsec$^2$)} &
\multicolumn{1}{c}{(arcsec)} &
\multicolumn{1}{c}{} &
\multicolumn{1}{c}{(mag/arcsec$^2$)} &
\multicolumn{1}{c}{(arcsec)} &
\multicolumn{1}{c}{(mag/arcsec$^2$)} &
\multicolumn{1}{c}{(arcsec)} &
\multicolumn{1}{c}{(arcsec)} &
\multicolumn{1}{c}{(mag/arcsec$^2$)} &
\multicolumn{1}{c}{(arcsec)} \\
\multicolumn{1}{c}{(1)} &
\multicolumn{1}{c}{(2)} &
\multicolumn{1}{c}{(3)} &
\multicolumn{1}{c}{(4)} &
\multicolumn{1}{c}{(5)} &
\multicolumn{1}{c}{(6)} &
\multicolumn{1}{c}{(7)} &
\multicolumn{1}{c}{(8)} &
\multicolumn{1}{c}{(9)} &
\multicolumn{1}{c}{(10)} &
\multicolumn{1}{c}{(11)} \\
\noalign{\smallskip}
\hline
\noalign{\smallskip}
  ESO 139$-$G009 & $17.30\pm0.05$ & $2.86\pm0.07$ & $1.35\pm0.03$ & $19.51\pm0.15$ & $15.12\pm1.08$ & $19.94\pm0.07$ & $14.40\pm0.10$ & $1.51\pm0.10$ & $21.49\pm0.11$ & $30.34\pm0.16$ \\
  IC 874       & $15.97\pm0.12$ & $2.40\pm0.15$ & $1.28\pm0.05$ & $18.59\pm0.08$ & $14.70\pm0.90$ & $17.62\pm0.02$ & $15.00\pm0.04$ & $3.53\pm0.05$ & $20.72\pm0.02$ & $38.61\pm0.04$ \\
  IC 4796      & $18.02\pm0.03$ & $3.61\pm0.02$ & $1.00\pm0.32$ & $19.70\pm0.04$ & $16.94\pm0.15$ & $20.47\pm0.02$ & $19.11\pm0.08$ & $1.56\pm0.07$ &                &                \\
  NGC 357      & $18.05\pm0.03$ & $3.93\pm0.03$ & $1.40\pm0.08$ & $21.63\pm0.21$ & $34.86\pm3.80$ & $20.23\pm0.07$ & $25.41\pm0.08$ & $1.64\pm0.05$ & $21.36\pm0.06$ & $60.06\pm1.37$ \\
  NGC 364      & $17.66\pm0.02$ & $2.39\pm0.01$ & $0.99\pm0.03$ & $19.22\pm0.02$ & $13.84\pm0.07$ & $19.41\pm0.02$ & $10.21\pm0.03$ & $1.48\pm0.02$ &                &                \\
  NGC 936      & $18.08\pm0.04$ & $8.36\pm0.18$ & $1.40\pm0.12$ & $20.42\pm0.28$ & $63.72\pm6.70$ & $19.86\pm0.05$ & $46.20\pm1.05$ & $6.62\pm0.05$ & $21.55\pm0.08$ &$113.85\pm2.57$ \\
  NGC 1308     & $18.05\pm0.05$ & $2.31\pm0.03$ & $1.15\pm0.23$ & $19.56\pm0.07$ & $14.39\pm0.07$ & $19.44\pm0.05$ & $14.17\pm1.32$ & $2.51\pm0.07$ &                &                \\
  NGC 1440     & $16.48\pm0.02$ & $3.65\pm0.08$ & $1.03\pm0.07$ & $18.31\pm0.13$ & $20.25\pm0.89$ & $18.37\pm0.02$ & $19.23\pm0.05$ & $4.13\pm0.05$ & $21.17\pm0.07$ & $50.35\pm0.32$ \\
  NGC 2950     & $18.48\pm0.07$ & $8.16\pm0.10$ & $3.06\pm0.53$ & $20.70\pm0.12$ & $35.61\pm0.95$ & $20.31\pm0.06$ & $32.20\pm0.83$ & $2.38\pm0.03$ &                &                \\
  NGC 3412     & $17.21\pm0.03$ & $3.83\pm0.06$ & $2.00\pm0.30$ & $18.98\pm0.13$ & $26.42\pm1.16$ & $19.13\pm0.03$ & $34.00\pm2.32$ &               & $22.24\pm0.04$ & $63.00\pm2.18$ \\
  NGC 3941     & $16.72\pm0.02$ & $4.24\pm0.02$ & $1.70\pm0.01$ & $17.43\pm0.02$ & $17.70\pm0.06$ & $19.49\pm0.03$ & $23.63\pm0.03$ & $0.96\pm0.03$ &                &                \\
  NGC 4340     & $17.60\pm0.03$ & $5.32\pm0.01$ & $1.27\pm0.01$ & $20.05\pm0.02$ & $33.38\pm0.15$ & $20.76\pm0.04$ & $61.51\pm0.06$ & $3.88\pm0.05$ &                &                \\
  NGC 6684     & $18.05\pm0.08$ & $8.77\pm0.20$ & $2.00\pm0.13$ & $18.14\pm0.05$ & $29.56\pm1.35$ & $19.58\pm0.07$ & $48.75\pm1.51$ & &    &                \\
  NGC 7079     & $18.51\pm0.01$ & $3.20\pm0.01$ & $0.90\pm0.04$ & $19.53\pm0.41$ & $14.96\pm1.72$ & $21.53\pm0.10$ & $18.95\pm0.08$ & $1.40\pm0.07$ & $22.29\pm0.15$ & $51.48\pm0.57$ \\
\noalign{\smallskip}
\hline
\noalign{\bigskip}
\end{tabular}
\begin{minipage}{23cm}
  NOTE.  Col.(2): effective surface brightness of bulge; Col.(3)
  effective radius of bulge; Col.(4): shape parameter of bulge;
  Col.(5): central surface brightness of disc; Col.(6): scale length
  of disc; Col.(7): central surface brightness of bar; Col.(8): bar
  length; Col.(9): scale length of flat bar; Col.(10): central surface
  brightness of lens; Col.(6): length of lens. 
\end{minipage}
\end{table}
\end{landscape}

\begin{figure*}[!htb]
\centering
\includegraphics[angle=0,width=15cm]{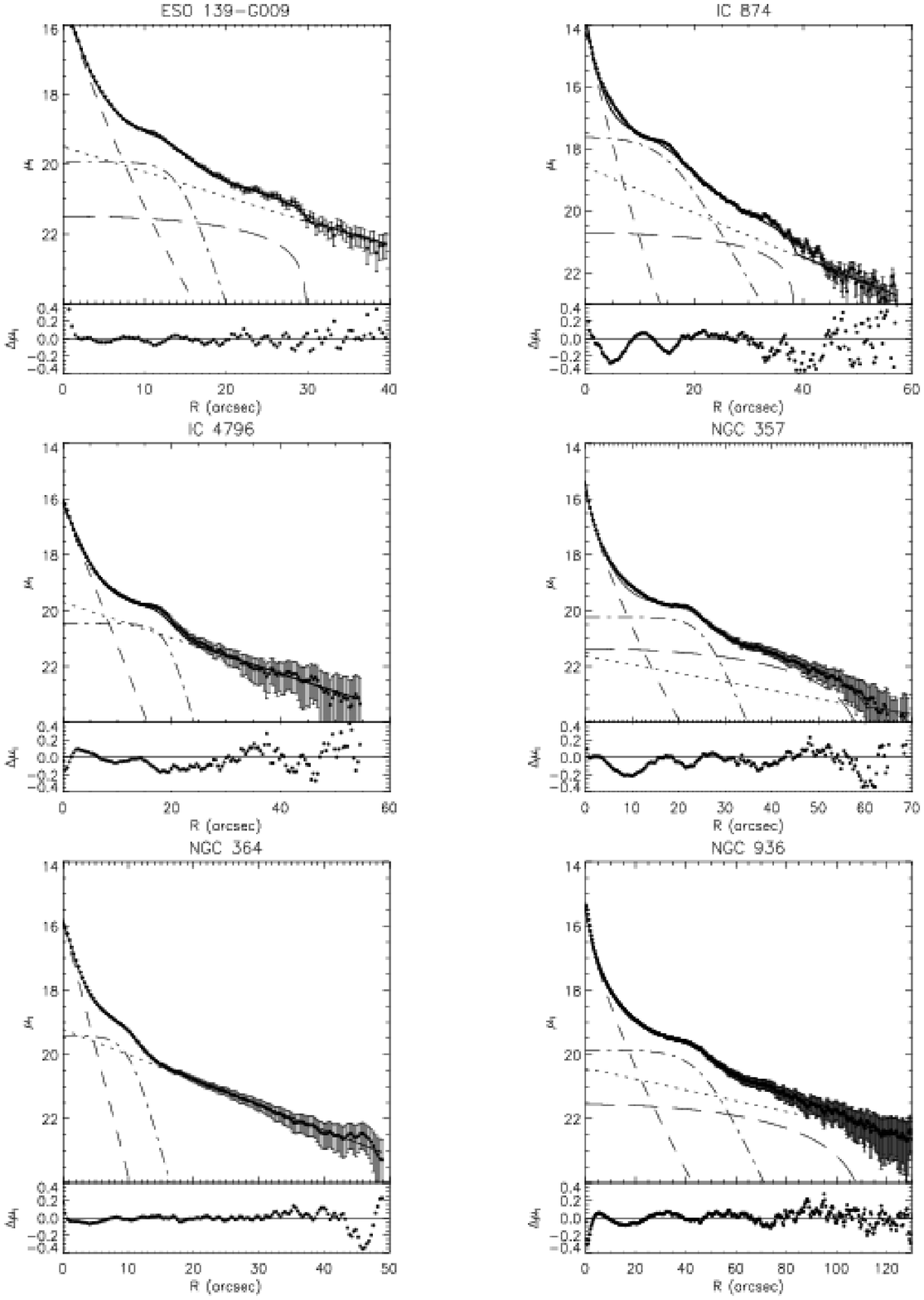}
\caption{Decomposition of the radial surface brightness profiles 
  in the $I$-band along the major axis of the bar of the sample galaxies,
  and residuals of the fit. The adopted structural components are
  bulges ({\it dashed line\/}), discs ({\it dotted line\/}), bar ({\it
  dash-dotted line\/}) and lenses ({\it long-dashed line\/}). The {\it
  continuous line\/} represents the total model.} 
\label{fig:decomposition}
\end{figure*}

\addtocounter{figure}{-1}

\begin{figure*}[!htb]
\centering
\includegraphics[angle=0,width=15cm]{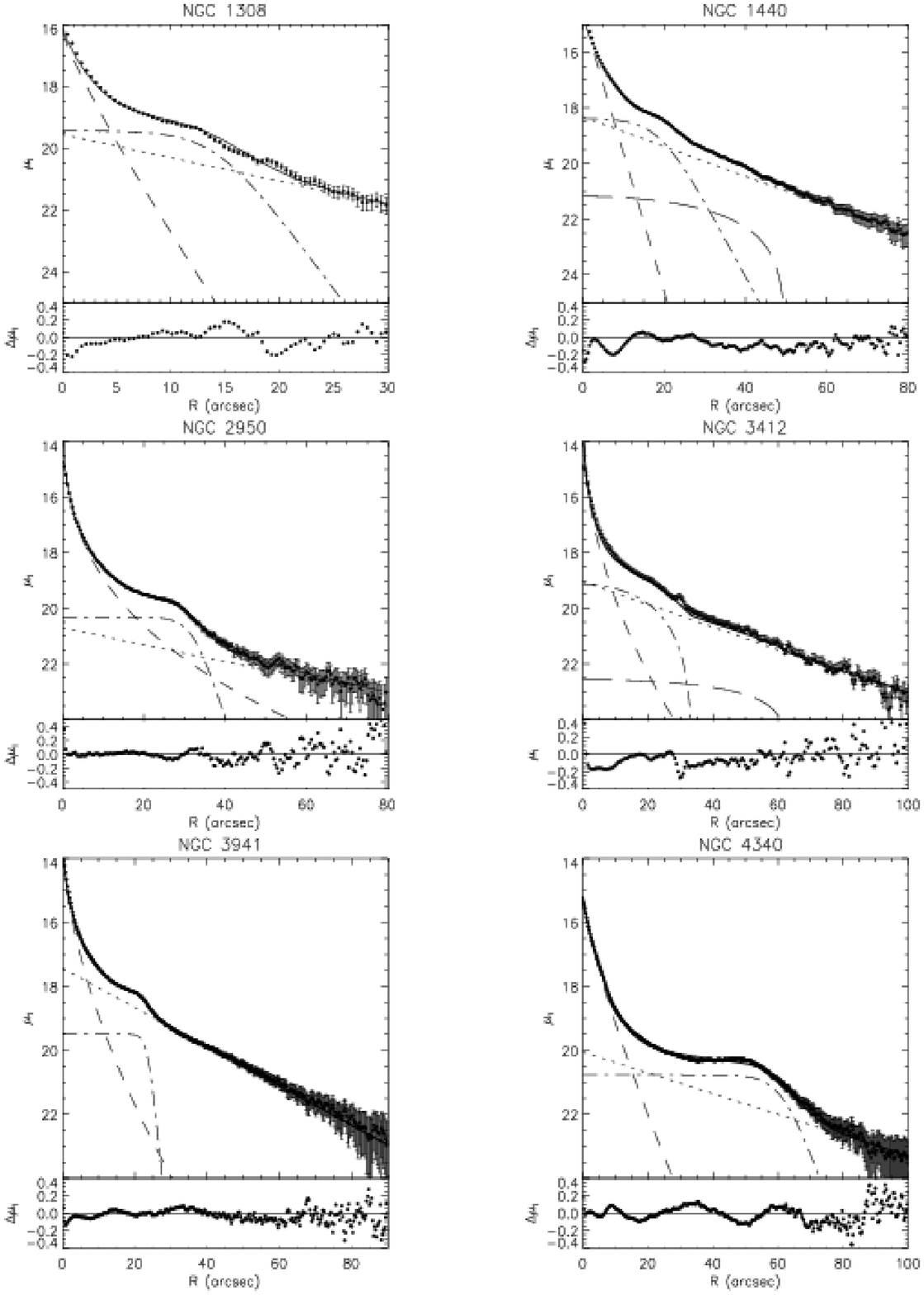}
\caption{(continued)}
\label{fig:decompositionb}
\end{figure*}

\addtocounter{figure}{-1}

\begin{figure*}[!htb]
\centering
\includegraphics[angle=0,width=15cm]{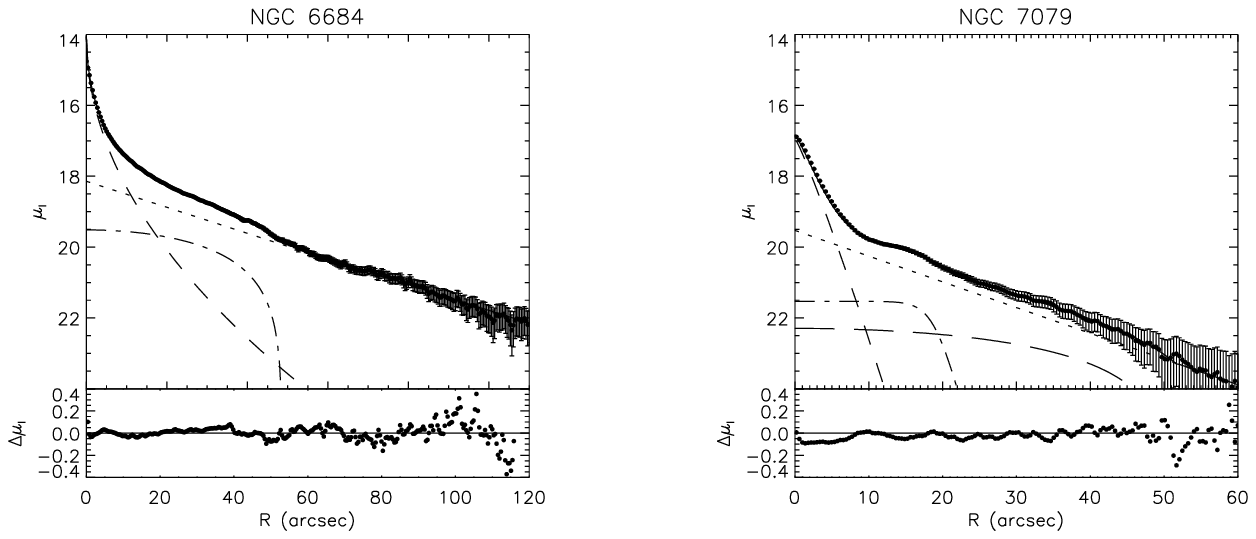}
\caption{(continued)}
\label{fig:decompositionc}
\end{figure*}

\subsection{Seeing effects}
\label{sec:decomposition_seeing}

The ground-based images are affected by seeing, which scatters the
light of the objects and produces a loss of spatial resolution. This
is particularly critical in the central regions of galaxies, where the
slope of the radial surface brightness profile is steeper. Since the
bulge contribution dominates the surface brightness distribution at
small radii, seeing mostly affects bulge structural parameters. Seeing
effects on the scale parameters of a S\'ersic surface brightness
profile have been extensively discussed by Trujillo et al.\ (2001a,b).

To deal with this limitation, we fitted the observed radial surface
brightness profiles at radii larger than twice the FWHM of the point
spread function (PSF). The PSF was approximated with a circular
two-dimensional Gaussian and its FWHM was derived by fitting several
field stars in the final combined images of the sample galaxies (Table
\ref{tab:log}).
To estimate systematic errors due to the seeing on the fitted
parameters of the bulge, we applied the adopted parametric
decomposition technique to a set of artificial disc galaxies.
We generated 250 images of galaxies with a S\'ersic bulge and an
exponential disc with surface brightnesses and scale lengths randomly
chosen in the range of the values that we observed for the sample
galaxies (Table \ref{tab:parameters}). This has been done by
satisfying all the following conditions: $2'' \leq r_{\rm e} \leq
10''$, $0.8\leq n\leq 3$, $12''\leq h
\leq 40''$, $0 \leq B/D \leq 0.5$, and $11\leq m_{I} \leq 14$.

The simulated images have been built to have the pixel scale, gain and
readout noise of the worst-resolved images of our dataset, namely
those obtained with the 1.54 m Danish telescope. An appropriate level
of noise was added to yield a signal-to-noise ratio similar to that of
the images we obtained during our observing runs.  Finally, all the
artificial images have been convolved with a circular two-dimensional
Gaussian of $\rm FWHM=1\farcs0$ which is the typical seeing FWHM of
our data (Table \ref{tab:log}).

The images of the artificial galaxies have been analyzed as if they
were real. The fitting algorithm recovered slightly fainter scale
surface brightnesses, larger scale lengths and larger bulge shape
parameters with respect to the input values. The mean relative errors
on the fitted parameters are $\left<\Delta \mu_{\rm e}\right>= -0.01$,
$\left<\Delta r_{\rm e}\right>= 0.02$, and $\left<\Delta
n\right>=0.02$ for the bulge, $\left<\Delta\mu_0\right>\lesssim-0.01$
and $\left<h\right>=0.01$ for the disc, providing the reliability and
accuracy of the bulge parameters derived.
We suspect that seeing affects bar parameters in the same way as it
does disc parameters. In fact, the radial surface brightness profile
of bars does not steepen toward the centre (Figure
\ref{fig:decomposition}) and the central surface brightnesses and
scale lengths of bars are similar to those of discs (Table
\ref{tab:parameters}).

\subsection{Dependence of the result on the fitting model}
\label{sec:decomposition_nolens}

The reason for including a bar and possibly a lens in fitting the
radial surface brightness profiles of SB0 galaxies is to reduce the
large and systematic residuals obtained when only the bulge and disc
are considered.

In order to investigate how the different structural parameters depend
on the fitting procedure, we have performed two other fits: a
photometric decomposition with only the bulge and disc components, and
a photometric decomposition with the bulge, disc and bar, but without
the lens component.
Table \ref{tab:nobar} lists the relative differences for the bulge and
disc parameters if only bulge and disc structures are used in fitting
the surface brightness profiles of the galaxies. In general, discs turn
out to be smaller and brighter than those obtained from fits that
include the bar and lens components. It should be noticed that the
sample galaxies host a strong bar, which is not a negligible light
component in the fitting procedure.
Table \ref{tab:nolens} lists the relative differences of the bulge,
disc and bar parameters if the lens is ignored in the fitting
procedure.  It turns out that ignoring the light contribution of the
lens does not affect the bar parameters since these are characterized
by negligible mean relative differences ($\left<\Delta \mu_{\rm
b}\right>=0.04$, $\left<\Delta r_{\rm b}\right>=0.04$).
This is in disagreement with the claims by Valenzuela \& Klypin
(2003), who argue that the bar length of the galaxies resulting from
their numerical simulations could hardly be compared to that of the
galaxies observed by Aguerri et al. (2003) because of the presence of
a lens component.

\begin{table}
\caption[]{Relative differences in the structural parameters if only bulge 
  and disc components are fitted to the surface brightness profiles of
  the galaxies.  They are defined as $\Delta
  X=\frac{X_{0}-X_{1}}{X_{0}}$ where $X_{0}$ is the parameter when all
  components are fitted and $X_{1}$ is the parameter when only bulge
  and disc components are used.}
\label{tab:nobar} 
\begin{tabular}{lrrrrr}
\hline
\noalign{\smallskip}
\multicolumn{1}{c}{Galaxy} &
\multicolumn{1}{c}{$\Delta \mu_{\rm e}$} &
\multicolumn{1}{c}{$\Delta r_{\rm e}$} &
\multicolumn{1}{c}{$\Delta n$} &
\multicolumn{1}{c}{$\Delta \mu_{0}$} &
\multicolumn{1}{c}{$\Delta h$} \\
\multicolumn{1}{c}{(1)} &
\multicolumn{1}{c}{(2)} &
\multicolumn{1}{c}{(3)} &
\multicolumn{1}{c}{(4)} &
\multicolumn{1}{c}{(5)} &
\multicolumn{1}{c}{(6)} \\
\noalign{\smallskip}
\hline
\noalign{\smallskip}
ESO 139$-$G009 & $-0.01$ & $ 0.41$ & $ 0.27$ & 0.08 & 0.58 \\
IC 874       & $-0.04$ & $ 0.01$ & $-0.18$ & 0.10 & 0.37\\
IC 4796      & $-0.01$ & $ 0.21$ & $-0.09$ & 0.06 & 0.37\\
NGC 357      & $-0.01$ & $ 0.18$ & $ 0.09$ & 0.14 & 0.58 \\
NGC 364      & $-0.03$ & $-0.38$ & $-0.40$ & 0.03 & 0.26 \\
NGC 936      & $-0.01$ & $ 0.13$ & $-0.23$ & 0.01 & 0.52 \\
NGC 1308     & $ 0.01$ & $ 0.44$ & $ 0.15$ & 0.09 & 0.59 \\
NGC 1440     & $-0.05$ & $-0.59$ & $-1.06$ & 0.05 & 0.26 \\
NGC 2950     & $ 0.06$ & $ 0.60$ & $ 0.34$ & 0.12 & 0.64 \\
NGC 3412     & $ 0.01$ & $ 0.31$ & $ 0.01$ & 0.06 & 0.55 \\
NGC 3941     & $ 0.02$ & $ 0.24$ & $ 0.13$ & 0.05 & 0.43 \\
NGC 4340     & $ 0.01$ & $ 0.18$ & $ 0.12$ & 0.07 & 0.31 \\
NGC 6684     & $ 0.03$ & $ 0.08$ & $ 0.10$ & 0.09 & 0.20 \\
NGC 7079     & $-0.18$ & $ 0.33$ & $ 0.17$ & 0.05 & 0.47 \\
\noalign{\smallskip}
\hline
\end{tabular}
\end{table}

\begin{table*}
\caption[]{Relative differences of the structural parameters for the galaxies 
  without a lens fitted to the surface brightness profiles  They are
  defined as $\Delta X=\frac{X_{0}-X_{1}}{X_{0}}$ where $X_{0}$ is the
  parameter when all components are fitted and $X_{1}$ is the
  parameter when the lens component is not included in the fit.}
\label{tab:nolens} 
\begin{tabular}{lrrrrrrr}
\hline
\noalign{\smallskip}
\multicolumn{1}{c}{Galaxy} &
\multicolumn{1}{c}{$\Delta \mu_{\rm e}$} &
\multicolumn{1}{c}{$\Delta r_{\rm e}$} &
\multicolumn{1}{c}{$\Delta n$} &
\multicolumn{1}{c}{$\Delta \mu_{0}$} &
\multicolumn{1}{c}{$\Delta h$} &
\multicolumn{1}{c}{$\Delta \mu_{\rm b}$} &
\multicolumn{1}{c}{$\Delta r_{\rm b}$} \\
\multicolumn{1}{c}{(1)} &
\multicolumn{1}{c}{(2)} &
\multicolumn{1}{c}{(3)} &
\multicolumn{1}{c}{(4)} &
\multicolumn{1}{c}{(5)} &
\multicolumn{1}{c}{(6)} &
\multicolumn{1}{c}{(7)} &
\multicolumn{1}{c}{(8)} \\
\noalign{\smallskip}
\hline
\noalign{\smallskip}
ESO 139$-$G009 & $-0.01$ & $ 0.07$ & $ 0.20$ & 0.05 & 0.27 & $-0.03$ & $ 0.04$\\
IC 874       & $-0.06$ & $-0.42$ & $-0.49$ & 0.05 & 0.19 & $-0.04$ & $-0.09$\\
NGC 357      & $-0.02$ & $-0.03$ & $-0.03$ & 0.11 & 0.49 & $-0.04$ & $-0.01$\\
NGC 936      & $-0.02$ & $-0.13$ & $-0.38$ & 0.07 & 0.41 & $-0.05$ & $-0.06$\\
NGC 1440     & $-0.03$ & $-0.39$ & $-0.58$ & 0.03 & 0.17 & $-0.08$ & $-0.13$\\
NGC 3412     & $-0.14$ & $-0.05$ & $ 0.05$ & 0.02 & 0.12 & $-0.02$ & $-0.06$\\
NGC 7079     & $-0.02$ & $ 0.07$ & $ 0.05$ & 0.04 & 0.21 & $ 0.01$ & $ 0.06$\\
\noalign{\smallskip}
\hline
\end{tabular}
\end{table*}

\section{Parameter relations}
\label{sec:discussion}

\subsection{Length and strength of the bar}

Measuring the bar length is not trivial task and several methods have
been developed for this purpose (see Athanassoula \& Misiriotis 2002
for a review).

After measuring the bar length by means of the photometric
decomposition ($r_{\rm b,1}$ in Table \ref{tab:bar}), we derived it by
means of Fourier decomposition of the galaxy surface brightness with
the methods of Aguerri et al.\ (2000) and Debattista \& Sellwood
(2000).

The method of Aguerri et al.\ (2000) is based on the ratios of the
intensities in the bar and the inter-bar region.  The azimuthal
surface brightness profile of the deprojected galaxies is decomposed
into a Fourier series.  The bar intensity, $I_{\rm b}$, then is
defined as $I_{\rm b}=I_{0}+I_{2}+I_{4}+I_{6}$ (where $I_{0},
I_{2},I_{4}$ and $I_{6}$ are the $m=0$, 2, 4 and 6 terms of the
Fourier decomposition, respectively).  The inter-bar intensity is
defined as $I_{\rm ib}=I_{0}-I_{2}+I_{4}-I_{6}$.  The bar region is
defined as the region where
\begin{equation}
I_{\rm b}/I_{\rm ib}>0.5[\max(I_{\rm b}/I_{\rm ib})-\min(I_{\rm
b}/I_{\rm ib})]+\min(I_{\rm b}/I_{\rm ib}).
\label{eq:bar_region}
\end{equation}
Thus the semi-major axis of the bar is identified as the outer radius
at which
\begin{equation}
I_{\rm b}/I_{\rm ib}=0.5[\max(I_{\rm b}/I_{\rm ib})-\min(I_{\rm
b}/I_{\rm ib})]+\min(I_{\rm b}/I_{\rm ib}).
\end{equation}
This is given as $r_{\rm b,2}$ in Table \ref{tab:bar} for all the
sample galaxies. According to the results by Athanassoula \&
Misiriotis (2002), based on numerical simulations, the error in bar
length obtained by the analysis of the Fourier amplitudes ranges
between $4\%$ and $8\%$.

The method of Debattista \& Sellwood (2000) is also based on a Fourier
decomposition of the surface brightness distribution. It assumes that
the bar can extend only in the radial interval where the phase of the $m=2$ moment is
constant to within the errors. Our sample galaxies were selected to be
free of spirals and rings to prevent the overestimation of the bar
length which is listed as $r_{\rm b,3}$ in Table
\ref{tab:bar}. 

A related, but not equivalent, method uses the phases of the ellipse
fits  (Debattista \& Williams 2004), which then gives our fourth
estimate of the bar length $r_{\rm b,4}$ in Table
\ref{tab:bar}.
The ellipses fitting the galaxy isophotes are deprojected and
their phase angle is measured. The bar length is then the largest
radius out to which the phases are consistent with a constant, taking
into account that deprojecting the bulge, which we did not subtract
from our image for this analysis, results in a twist interior to the
bar.

The mean of $r_{\rm b,1}$, $r_{\rm b,2}$, $r_{\rm b,3}$ and $r_{\rm
b,4}$ is our best estimate of the bar length $r_{\rm b}$ and we used
the largest deviations from the mean for the error estimates (Table
\ref{tab:bar}).

The bar strength $s_{\rm b}$ of each sample galaxies is also listed
Table \ref{tab:bar} and is defined as
\begin{equation}
s_{\rm b}=\frac{1}{r_{\rm b,2}-r_{\rm in}} \int^{r_{\rm b,2}}_{r_{\rm
in}} \frac{I_{2}}{I_{0}}(r) dr,
\end{equation}
where $r_{\rm in}$ and $r_{\rm b,2}$ are the inner and outer radius of
the bar region defined by Eq. \ref{eq:bar_region}, while $I_0$ and
$I_2$ are the $m=0$ and $m=2$ terms of the Fourier decomposition,
respectively.
Although ESO 139$-$G009 is classified as a weakly barred S0 (RC3), its
bar strength is comparable to that of the other SB0 galaxies in the
sample. 

We found a mean bar-to-disc scale length ratio $\left< r_{\rm
b}/h \right>=1.21\pm0.08$. This becomes $\left< r_{\rm b}/h
\right>=1.42\pm0.09$ if lenses are excluded from the surface
brightness fit.
This is in agreement with the findings of the recent numerical
simulations of barred galaxies by O'Neil \& Dubinski (2003, $\left<
r_{\rm b}/h \right>= 1.11\pm0.33$). On the contrary, the bars in the
numerical simulations by Valenzuela \& Klypin (2003) have a smaller
bar-to-disk ratio ($\left< r_{\rm b}/h \right> =1.0$).

\begin{table*}
\begin{center}
\caption[]{Length and strength of the bars in the sample galaxies }
\label{tab:bar} 
\begin{tabular}{lccccccc}
\hline
\noalign{\smallskip}
\multicolumn{1}{c}{Galaxy} &
\multicolumn{1}{c}{$r_{\rm b,1}$} &
\multicolumn{1}{c}{$r_{\rm b,2}$} &
\multicolumn{1}{c}{$r_{\rm b,3}$} &
\multicolumn{1}{c}{$r_{\rm b,4}$} &
\multicolumn{1}{c}{$r_{\rm b}$} &
\multicolumn{1}{c}{Ref.} &
\multicolumn{1}{c}{$s_b$} \\
\multicolumn{1}{c}{} &
\multicolumn{1}{c}{(arcsec)} &
\multicolumn{1}{c}{(arcsec)} &
\multicolumn{1}{c}{(arcsec)} &
\multicolumn{1}{c}{(arcsec)} &
\multicolumn{1}{c}{(arcsec)} &
\multicolumn{1}{c}{} &
\multicolumn{1}{c}{} \\
\multicolumn{1}{c}{(1)} &
\multicolumn{1}{c}{(2)} &
\multicolumn{1}{c}{(3)} &
\multicolumn{1}{c}{(4)} &
\multicolumn{1}{c}{(5)} &
\multicolumn{1}{c}{(6)} &
\multicolumn{1}{c}{(7)} &
\multicolumn{1}{c}{(8)} \\
\noalign{\smallskip}
\hline
\noalign{\smallskip}
ESO 139$-$G009 & 14.4 & 23.4 & 14.0 & 16.1 & $17.0^{+6.4}_{-3.0}$ & 1 & $0.30\pm0.03$ \\
IC 874       & 15.0 & 18.7 & 21.0 & 25.0 & $19.9^{+5.1}_{-4.9}$ & 1 & $0.52\pm0.02$ \\
IC 4796      & 19.1 & 20.8 & 20.0 & 23.0 & $20.7^{+2.3}_{-1.6}$ & 2 & $0.38\pm0.03$ \\
NGC 357      & 25.4 & 28.9 & 30.9 & 31.0 & $29.1^{+1.9}_{-3.7}$ & 2 & $0.44\pm0.02$ \\
NGC 364      & 10.2 & 13.3 & 15.5 & 17.5 & $14.1^{+3.4}_{-3.9}$ & 2 & $0.40\pm0.02$ \\
NGC 936      & 46.2 & 50.0 & 46.5 & 69.0 & $53.0^{+16}_{-6.8}$  & 2 & $0.42\pm0.01$ \\
NGC 1308     & 14.2 & 13.2 &  9.0 & 13.2 & $12.4^{+1.8}_{-3.4}$ & 1 & $0.44\pm0.02$ \\
NGC 1440     & 19.2 & 23.2 & 24.6 & 30.5 & $24.4^{+6.1}_{-5.2}$ & 1 & $0.37\pm0.04$ \\
NGC 2950     & 32.2 & 36.5 & 35.6 & 33.7 & $34.5^{+2.0}_{-2.3}$ & 2 & $0.40\pm0.03$ \\
NGC 3412     & 34.0 & 28.2 & 30.0 & 32.2 & $31.1^{+2.9}_{-2.9}$ & 1 & $0.34\pm0.02$ \\
NGC 3941     & 23.6 & 25.5 & 29.0 & 32.0 & $27.5^{+4.5}_{-3.9}$ & 2 & $0.29\pm0.03$ \\
NGC 4340     & 61.5 & 58.0 & 40.5 & 64.0 & $56.0^{+8.0}_{-15.5}$ & 2 & $0.49\pm0.01$ \\
NGC 6684     & 48.8 & 46.0 & 40.0 & 42.0 & $44.2^{+4.6}_{-4.2}$ & 2 & $0.42\pm0.03$ \\
NGC 7079     & 19.0 & 28.9 & 25.0 & 21.9 & $23.7^{+5.2}_{-4.7}$ & 3 & $0.22\pm0.06$ \\
\noalign{\smallskip}
\hline
\noalign{\bigskip}
\end{tabular}
\begin{minipage}{17cm}
  NOTE.  Col. (2): bar length from photometric decomposition;
  Col. (3): bar length from the ratio of the intensities in the bar
  and inter-bar region; Col. (4): bar length from $m=2$ moment phase;
  Col. (5): bar length from phase of the ellipse fit; Col. (6): mean
  value of the bar length; Col. (7): list of references for the bar
  length from Fourier analysis: 1=Aguerri et al.\ (2003), 2=this
  paper, 3=Debattista \& Williams (2004); Col. (8): bar strength from
  Fourier analysis. 
\end{minipage}
\end{center}
\end{table*}

\subsection{Bulge and disc interplay}

From NIR observations of spiral galaxies, Andredakis et al.\ (1995)
found a correlation between the bulge shape parameter and
morphological type. The bulges of early-type spirals are characterized
by $n\approx4$ (i.e., they have a de Vaucouleurs radial surface
brightness profile) while the bulges of late-type spirals are
characterized by $n\approx1$ (i.e., they have an exponential radial
surface brightness profile).  This result has been confirmed in
 various studies (e.g., de Jong 1996, MacArthur et al.\ 2003, and M\"ollenhoff 2004).

However, Balcells et al. (2003) performed bulge-disc decomposition of
combined space- and ground-based profiles of a sample of S0s and
early-type unbarred spirals and found that their bulges have a smaller
S\'ersic shape parameter than thought before
($\left<n\right>=1.7\pm0.7$). They claim that the high $n$ values of
ground-based profiles are the result of unresolved nuclear
components blending with the bulge light because of seeing.
We modeled the surface brightness distribution of the sample galaxies
by excluding the central region of the light profiles and therefore
our results are  not sensitive to this effect. Our sample bulges have a mean shape
parameter $\left<n\right>=1.45\pm0.15$ ($\left<n\right>=1.56\pm0.16$
if lenses are excluded from the fit).
This means that bulges of our SB0 galaxies have almost exponential
radial surface brightness profiles which are very similar to those
found by Balcells et al. (2003). 

\begin{figure}
\centering
\includegraphics[angle=0,width=9cm]{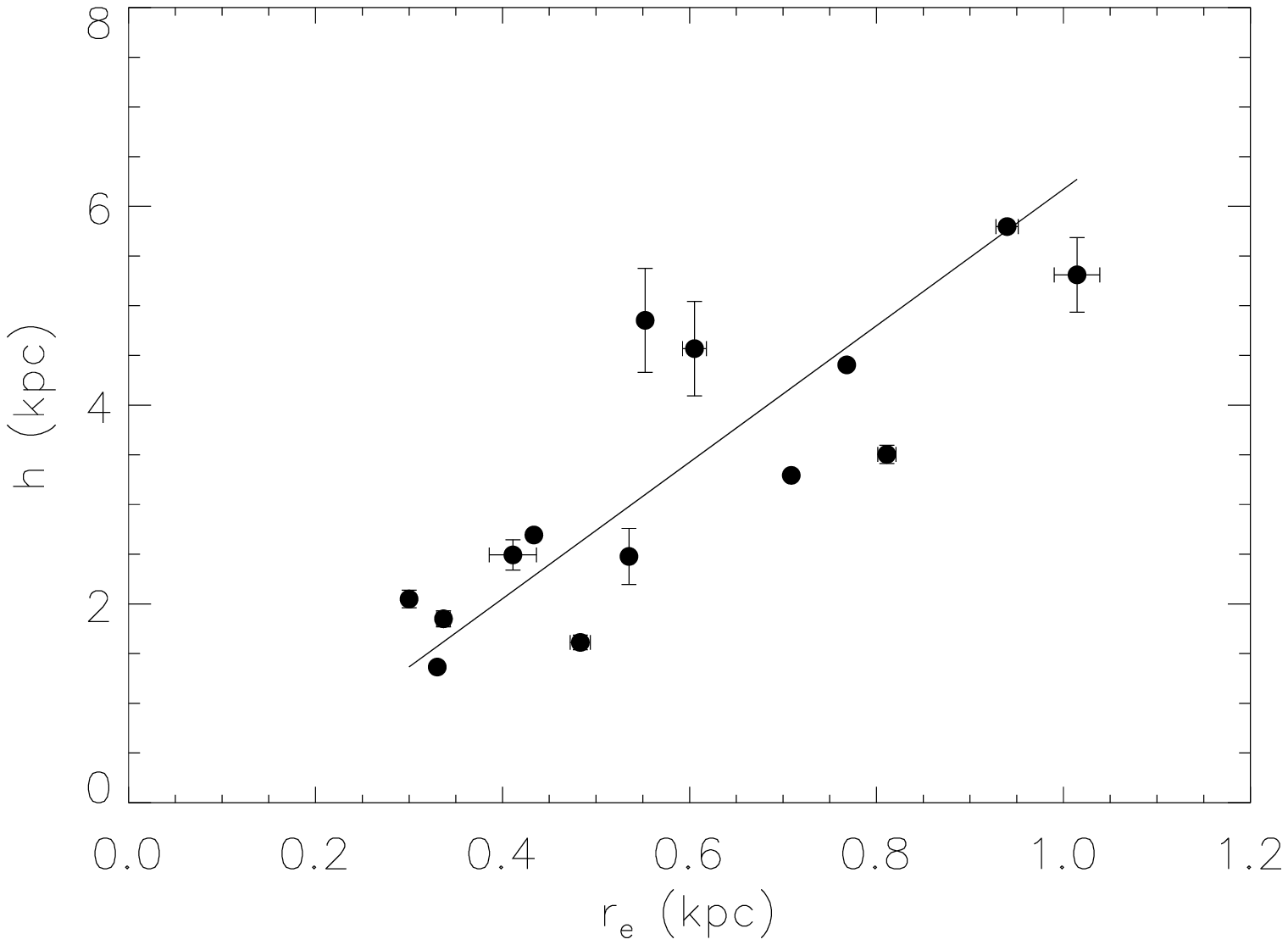}
\caption{Disc scale length versus bulge effective radius  
  for the sample galaxies. }
\label{fig:scalelengths}
\end{figure}

We found a strong correlation between the bulge effective radius and
the disc scale length in our sample galaxies (Pearson coefficient
$r=0.91$, Figure \ref{fig:scalelengths}). The mean value varies from
$\left<r_{e}/h\right>=0.20\pm0.01$ to
$\left<r_{e}/h\right>=0.22\pm0.02$, depending on whether the lens
component is or is not taken into account in the surface brightness
fit.
These values cannot be compared with the $I-$band ratios found by
Graham \& Prieto (1999), MacArthur et al. (2003) and  M\"ollenhoff
(2004) for mostly unbarred galaxies, because these recent
studies include only few or no early-type spirals and no S0
galaxies. Nevertheless, the $r_{e}/h$ found for our SB0 galaxies
favors the idea that this ratio may be only mildly (MacArthur et
al. 2003) and not strongly (Graham \& Prieto 1999; M\"ollenhoff 2004)
correlated with Hubble type, and it possibly increases from late- to
early-type disk galaxies.

 Since $\left<M_{\rm bulge}\right>=-24.86\pm1.33$ ($\left<M_{\rm
bulge}\right>=-24.69\pm1.11$ if the lenses are excluded from the fits)
we can conclude that our bulges are significantly brighter than the
few early-type spiral galaxies studied by M\"ollenhoff (2004).

\subsection{The location of SB0 bulges on the fundamental plane}
\label{sec:fp}

We collected from the literature the measurements of the central
stellar velocity dispersion of all our sample galaxies (Prugniel 
\& Simien 1997; Fisher 1997; Aguerri et al.\ 2003; Corsini et al.\ 2003;
Wegner et al.\ 2003), except for NGC 364.
For comparison, we corrected the kinematic data to the equivalent
of an aperture of radius $r_{\rm e}/8$ following the prescription by
J\/orgensen et al.\ (1995), where $r_{\rm e}$ is the effective radius. The
aperture-corrected velocity dispersions are given in Table
\ref{tab:sample}. We adopted the structural parameters derived in Sect.
\ref{sec:decomposition_lens} and given in Table 
\ref{tab:parameters}.

We used the Coma galaxies studied by Scodeggio et al.\ (1998) to
obtain a fundamental plane (hereafter FP) template relation for the
unbarred early-type galaxies in the $I$ band. For this purpose we
selected only the elliptical galaxies and the bulges of the unbarred
S0 galaxies, and we applied the aperture-correction to $r_{\rm e}/8$
to their central velocity dispersions. The parameters describing the
template FP were derived by assuming the Coma cluster at rest in the
CMB reference and $H_0 = 75$ km s$^{-1}$ Mpc$^{-1}$ with the result
that
\begin{equation}
\log r_{\rm e} = 1.12(\pm0.10) \log \sigma_0 + 0.33(\pm0.05) \mu_{\rm e} -8.68(\pm0.08) 
\end{equation}                                                    
\noindent
by minimizing the  square root of the residuals along the $\log
r_{\rm e}$ axis. No statistically significant difference was observed
when only elliptical galaxies or only bulges of unbarred S0 galaxies
are considered. The dispersion around this relation $\sigma=0.098$ and
has been measured as the rms scatter in the residuals of $\log r_{\rm
e}$.

\begin{figure}
\centering
\includegraphics[angle=0,width=9cm]{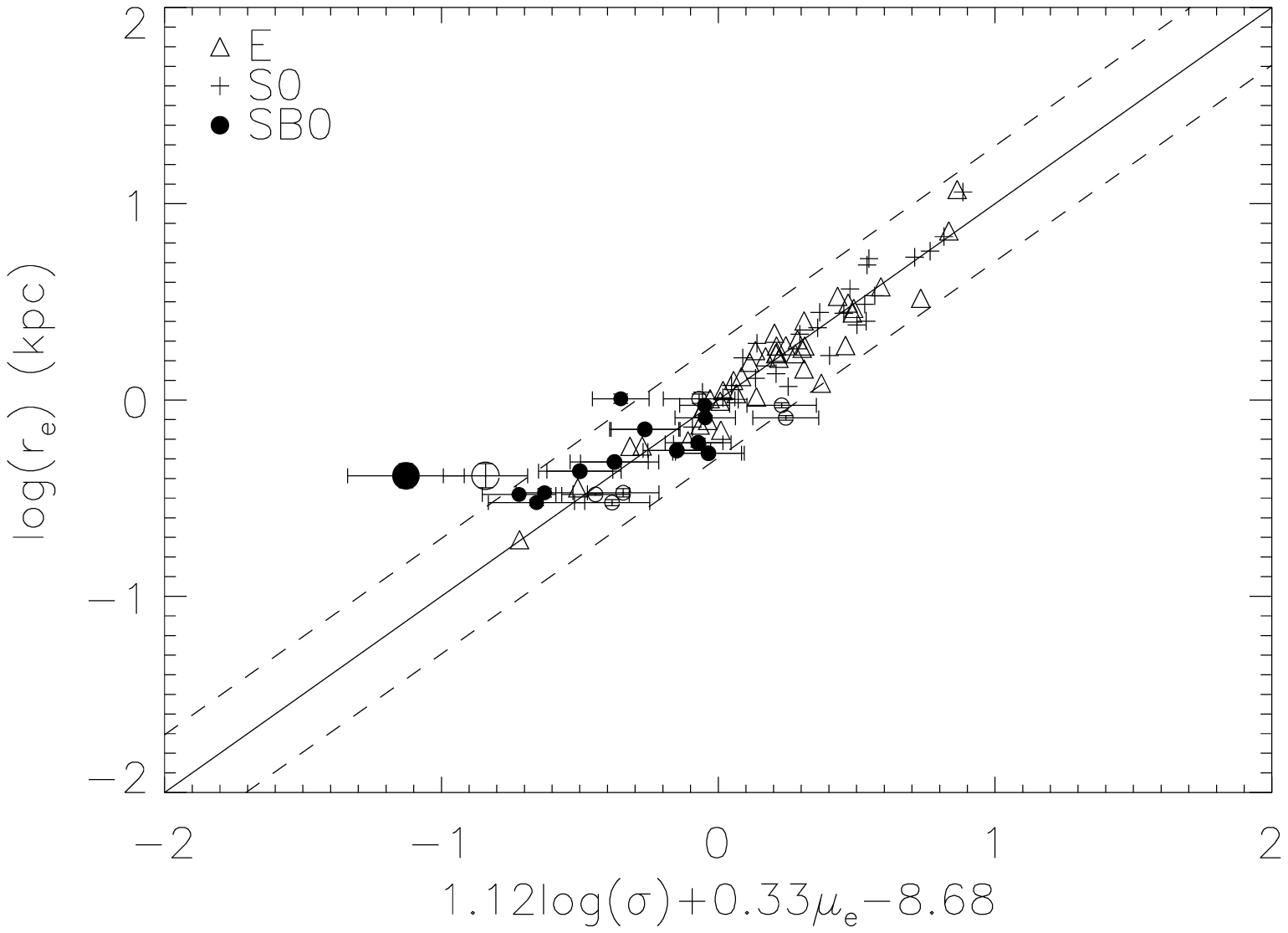}
\caption{Edge-on view of the fundamental plane ({\it continuous
    line\/}) of the Coma elliptical galaxies ({\it triangles\/}) and
    S0 bulges ({\it open triangles\/}) of Scodeggio et al. (1998). SB0
    bulges are indicated by {\it filled circles\/} when rotation is not
    taken into account to estimate the kinetic energy and by {\it
    open circles\/} when rotation is included. The dashed lines
    represent the 3$\sigma$ deviation from the FP.  The largest
    symbol corresponds to IC 874.}
\label{fig:fp}
\end{figure}

Figure \ref{fig:fp} shows an edge-on view of the template FP, where we
overplotted our SB0 bulges. The bulge of IC 874 is located outside the
template FP by more than $3\sigma$. We find that this is true even
adopting the structural parameters derived in
Sect. \ref{sec:decomposition_nolens} by means of a photometric
decomposition that does not include the lens component.
It should be noted that the location of bulges on the FP does not
depend on the choice of a S\'ersic law instead of a de Vaucouleurs law
for their radial surface brightness profile (Falc\'on-Barroso et al.\
2002).

The FP is related to the virial theorem, which is a consequence of the
balance of the kinetic and potential energies of systems in
equilibrium.  Giant elliptical galaxies do show any significant
rotation. The mean stellar rotation is therefore not taken into
account in estimating their kinetic energy.  
 This is not the case for fainter elliptical galaxies (Davies et
al. 1983) and bulges (Kormendy 1982; Kormendy \& Illingworth 1982)
which rotate as rapidly as models of isotropic oblate spheroids
(Binney 1978).

Following Prugniel \& Simien (1994), we have investigated the
contribution of rotation to the total kinetic energy  for the nine
SB0 bulges for which the stellar velocity curve has been measured.
This contribution is given by 
\begin{equation}
T \propto \left<V^{2}\right> + 3\left<\sigma^{2}\right>,
\end{equation}
assuming an isotropic tensor of velocity dispersions.
For isotropic oblate models the square maximum rotation velocity,
$V_{\rm max}^2$, and square central velocity dispersion, $\sigma_0^2$,
are a good approximation of the mass-weighted mean-square rotational
velocity, $\left< V^2 \right>$ and  velocity dispersion, $\left<
\sigma^2 \right>$, respectively (Binney 1980). 

 We measured the maximum rotation velocity $V_{\rm max}$ and
ellipticity $\epsilon$ of the bulge from the stellar velocity curve
and radial ellipticity profile at a radial distance corresponding to
$r_{\rm e}$, respectively.
For each galaxy we derived the ratio $(V_{\rm max}/\sigma_0)^\ast$ of
$V_{\rm max}/\sigma_0$ to the value predicted for the observed
ellipticity by the oblate models (Table \ref{tab:vsigma}). Our SB0
bulges have a weighted mean $\left< (V_{\rm max}/\sigma_0)^\ast
\right> = 0.94 \pm 0.21$ which is consistent with the early results of
Kormendy (1982). Following Kormendy (1982) we conclude that most
SB0 bulges rotate as fast as both bulges of unbarred galaxies and
oblate spheroid models. This is not the case of the bulges of NGC 1308
and NGC 4340 which are rotating significantly faster than oblate
spheroid models and are more similar to the triaxial SB0 bulges
studied by Kormendy (1982). The values of $V_{\rm max}/\sigma_0$ and
$(V_{\rm max}/\sigma_0)^\ast$ we obtained for NGC 936, NGC 2950 and
NGC 4340 can be compared to those obtained by Kormendy (1982). The
agreement is good for NGC 2950 but not for NGC 936 and NGC 4340.  We
attribute this discrepancy to the different estimate we adopted for
the $V_{\rm max}$ value.

\begin{table*}
\begin{center}
\caption[]{Structural parameters of the sample galaxies}
\label{tab:vsigma} \begin{tabular}{lccccc}
\hline
\noalign{\smallskip}
\multicolumn{1}{c}{Galaxy} &
\multicolumn{1}{c}{$V_{\rm max}$} &
\multicolumn{1}{c}{$V_{\rm max}/\sigma_0$} &
\multicolumn{1}{c}{$(V_{\rm max}/\sigma_0)^\ast$} &
\multicolumn{1}{c}{$\epsilon$} &
\multicolumn{1}{c}{Ref.} \\
\multicolumn{1}{c}{ } &
\multicolumn{1}{c}{(km s$^{-1}$)} &
\multicolumn{1}{c}{} &
\multicolumn{1}{c}{} &
\multicolumn{1}{c}{} &
\multicolumn{1}{c}{} \\
\multicolumn{1}{c}{(1)} &
\multicolumn{1}{c}{(2)} &
\multicolumn{1}{c}{(3)} &
\multicolumn{1}{c}{(4)} &
\multicolumn{1}{c}{(5)} &
\multicolumn{1}{c}{(6)} \\
\noalign{\smallskip}
\hline
\noalign{\smallskip}
ESO 139$-$G009 &$88\pm18$ & $0.45\pm0.13$ & $1.16\pm0.33$ & $0.13\pm0.02$ & 1\\
       IC 874 &  $44\pm3$ & $0.45\pm0.20$ & $1.77\pm0.79$ & $0.06\pm0.01$ & 1\\
     NGC  936 & $78\pm10$ & $0.38\pm0.05$ & $1.03\pm0.13$ & $0.12\pm0.01$ & 2\\
    NGC 1308 & $71\pm10$ & $0.32\pm0.09$ & $1.41\pm0.37$ & $0.05\pm0.01$ & 1\\ 
    NGC 1440 &  $68\pm6$ & $0.35\pm0.06$ & $1.05\pm0.17$ & $0.10\pm0.03$ & 1\\
     NGC 3412 &  $31\pm6$ & $0.28\pm0.18$ & $0.63\pm0.40$ & $0.16\pm0.05$ & 1\\
     NGC 2950 &  $93\pm7$ & $0.57\pm0.04$ & $0.87\pm0.07$ & $0.30\pm0.03$ & 3\\
     NGC 3941 & $45\pm11$ & $0.33\pm0.08$ & $0.81\pm0.20$ & $0.14\pm0.03$ & 4\\
     NGC 4340 &  $35\pm9$ & $0.30\pm0.08$ & $1.71\pm0.43$ & $0.03\pm0.01$ & 5\\
\noalign{\smallskip}
\hline
\noalign{\bigskip}
\end{tabular}
\begin{minipage}{17cm}
NOTE.  Col. (2): bulge velocity measured at $r_e$ given by Table
\ref{tab:parameters}; Col. (3): ratio of the bulge velocity to 
central dispersion given in Table \ref{tab:sample}; Col. (4): ratio
of $V_{\rm max}/\sigma_0$ to the value predicted for the observed
ellipticity by the oblate models (see Kormendy \& Illingworth 1982);
Col. (5): ellipticity of the bulge $\epsilon=1-b/a$ at $r_e$;
Col. (6): list of references for the stellar velocity curve: 1=Aguerri
et al.\ (2003), 2=Kormendy (1983), 3=Corsini et al. (2003),
4=Fisher (1997), 5=Kormendy (1982).
\end{minipage}
\end{center}
\end{table*}

By including the contribution of rotation, the scatter of SB0 bulges
with respect to the template FP is reduced by about $20\%$. This is in
agreement with results of Prugniel \& Simien (1994). However, the
bulge of IC 874 remains located outside the template FP (Figure
\ref{fig:fp}).

We discount the possibility that this effect is due to either a
disturbed morphology or tidal interaction with nearby companions.
IC 874 has been selected to have an undisturbed morphology (see
Sect. \ref{sec:sample}). This is confirmed by the analysis of the
surface photometry obtained in Sect. \ref{sec:decomposition}.
IC 874 has five nearby companions within a projected distance of 500 kpc
and a systemic velocity difference of less than 500 km s$^{-1}$ (e.g.,
Aguerri 1999).
We have computed the tidal radius for each of them, finding that
shortest one corresponds to ESO 508$-$G039 and  is about 37 kpc.
Since the optical radius of IC 874 is about 5 kpc (from $D_{25}/2$ in
Table \ref{tab:sample}) we conclude it does not suffer any tidal
interaction.
 Finally, the location of IC 874 outside of the template FP does
not depend on the inclusion  of the lens component in the fit of
the surface brightness profile.

\subsection{The location of SB0 bulges in the Faber--Jackson relation}
\label{sec:fj}

One of the projections of the FP is the Faber--Jackson relation
(hereafter FJ) which  relates the luminosity of elliptical
galaxies and S0 bulges to their central velocity dispersion (Faber
\& Jackson 1976).

\begin{figure}
\centering
\includegraphics[angle=0,width=9cm]{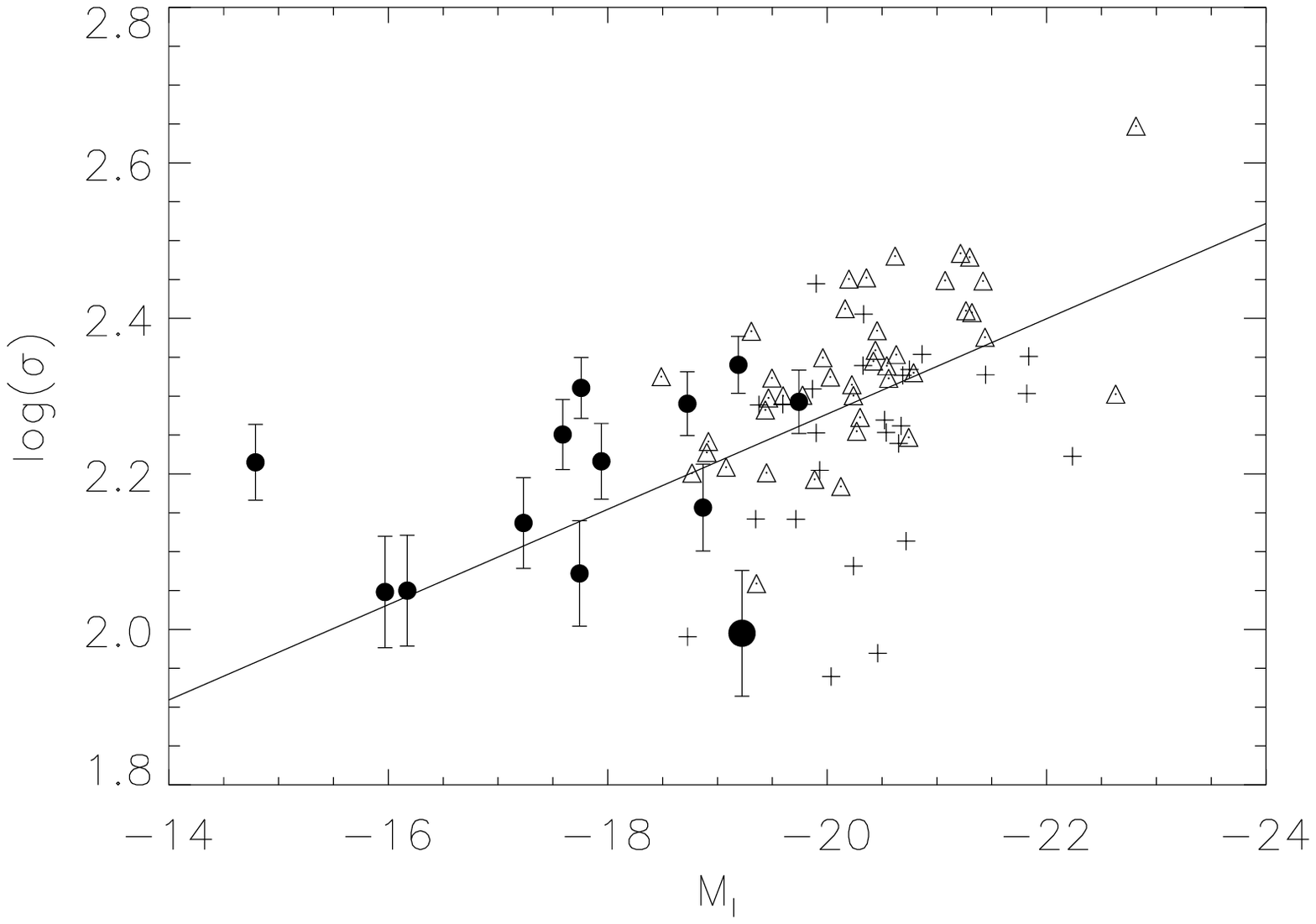}
\caption{Ellipticals (triangles) and bulges of S0 (crosses) galaxies
  from the Coma cluster. The full line represents the FJ relation for
  bulges of S0 and E galaxies. The full points correspond to the bulges of
  SB0 galaxies.  The largest symbol corresponds to IC 874.}

\label{fig:fj}
\end{figure}

SB0 bulges are consistent with the template FJ relation we built using
the same sample of Coma ellipticals and S0 bulges which has been
adopted in Sect. \ref{sec:fp} to obtain the template FP (Figure
\ref{fig:fj}). Nonetheless, the bulge of IC 874 is a low-$\sigma$
outlier and falls in the region of late-type bulges, which are usually
characterized by a lower velocity dispersion or equivalently a higher
luminosity with respect to their early-type counterparts (Kormendy \&
Kennicutt 2004).  This does not depend on whether the lens is included or
not in the fit of the surface brightness profile of the galaxy.

\section{Discussion and conclusions}
\label{sec:conclusions}

The formation of bulges is still an open question and several
mechanisms have been proposed (see Wyse et al. 1997, for a review). It
has been claimed that bulges formed before discs via hierarchical
clustering merging processes (Kauffmann \& White 1993), at the same
time as discs during a ``monolithic'' collapse process (Eggen et al.\
1962) or after discs as a results of a secular evolution processes
(Kormendy 1979). Furthermore, bulges can experience acquisition events
via infall of external material or accretion of satellite galaxies
(e.g., Aguerri et al.\ 2001). Up to now none of these pictures has
been  able
to reproduce all the observed properties of bulges along the Hubble
sequence of disc galaxies, leading to the idea that several of the
processes outlined in the different scenarios could have played a role
in forming a galaxy (Bouwens et al.\ 1999).

One issue that is not well understood is the influence of bars on
bulge formation. Bars are very efficient mechanisms for driving
material from the outermost to the innermost regions of spiral
galaxies, and produce the formation and/or growth of the bulge
component.
Recently, Debattista et al.\ (2004) have studied the bar-buckling
instability through a set of high resolution collisionless $N$-body
simulations. They conclude that the bar formation drives material to
central regions of the disc, building a component which shows an
almost exponential surface brightness profile.
 Although we observed that the SB0 bulges have almost exponential
surface brightness profiles, it is difficult to conclude that their
assembly is part of the bar formation process. In fact, SB0 bulges
have the same shape parameters as bulges of early-type unbarred
galaxies (Balcells et al. 2003). This means that it is difficult to
photometrically distinguish the influence of the bar on the formation
of SB0 bulges. Nevertheless, the fact that bulges of SB0 galaxies have
almost exponential surface brightness profiles suggest that they were
not formed by violent relaxation in mergers.

In a recent review about the secular evolution of galactic discs,
Kormendy \& Kennicutt (2004) discussed the properties of pseudobulges,
which are the dense central components in some lenticular and spiral
galaxies. In contrast to classical merger-built bulges, pseudobulges
are disc-like structures made slowly out of disc gas.
 We find that the bulges of IC 874, NGC 1380 and NGC 4340 are the most
reliable pseudobulges in our sample. The bulges of NGC 1308 and NGC
4340 are  more
rotation-dominated than classical bulges and oblate spheroid models in
the $V_{\rm max}/\sigma_{0}$--$\epsilon$ diagram (e.g., Kormendy \&
Kennicutt 2004). The bulge of IC 874 is a low-$\sigma$ outlier in the
FJ relation and it is outside the $3\sigma$ band in the FP of
elliptical galaxies and S0 bulges. Since IC 874, NGC 1308 and NGC 4340
host the
strongest bars of the sample (Table \ref{tab:bar}) we could speculate
that the formation of their bulges has been influenced by the bar.
We conclude that the rest of the sample galaxies have classical bulges,
which follow the FP and FJ relations traced by bulges of unbarred S0
and elliptical galaxies and rotate as fast as bulges of
unbarred galaxies.

\begin{acknowledgements}

We would like to thank the anonymous referee for detailed comments
which improved the outline and content of the paper.
We are most grateful to Victor P. Debattista for making available
his $I$-band imaging of NGC 7079 for the present work and for his help
in measuring bar length.
We acknowledge Luca Ciotti, Cesar Gonz\'alez-Garc\'{\i}a, and John
Kormendy for useful discussions.
EMC thanks the Instituto de Astrof\'\i sica de Canarias for the
hospitality while this paper was in progress. JALA, NER and CMT have
been founded by the Spanish DGES, grant AYA2001-3939.
This research has made use of the Lyon Meudon Extragalactic data base
(LEDA) and of the NASA/IPAC Extragalactic data base (NED). This paper
is based on observations carried out with the New Technology Telescope
and the Danish 1.54 m Telescope (Prop. No. 67.B-0230, 68.B-0329,
69.B-0706, and 70.B-0338) at the European Southern Observatory in La
Silla (Chile), with the Italian Telescopio Nazionale Galileo
(Prop. AOT-7, TAC\_25) operated on the island of La Palma by the
Centro Galileo Galilei of the Istituto Nazionale di Astrofisica, and
with the Jacobus Kapteyn Telescope operated by the Isaac Newton group
on La Palma at the Spanish del Roque de los Muchachos Observatory of
the Instituto de Astrof\'\i sica de Canarias.

\end{acknowledgements}

\end{document}